\documentclass[11pt, oneside]{article}   	
\usepackage{geometry}                		
\usepackage{graphicx}				
\usepackage{amssymb}
\usepackage{amsmath}
\usepackage{natbib}

\usepackage{hyperref}
\newcommand{\footremember}[2]{%
    \footnote{#2}
    \newcounter{#1}
    \setcounter{#1}{\value{footnote}}%
}
\newcommand{\footrecall}[1]{%
    \footnotemark[\value{#1}]%
} 

\graphicspath{{images/}}

\title{On the measurement of the dispersion relation by a radar and the implication on the current retrieval}
\author{Susanne St{\o}le-Hentschel 
\footremember{TAU}{School of Engineering, Tel Aviv University, Israel}
\footremember{ENS}{Universit'e Paris-Saclay, ENS Paris-Saclay, Centre Borelli, Gif-sur-Yvettes, France}
Benjamin Keeler Smeltzer
\footremember{NTNU}{Department of Energy and Process Engineering, Norwegian University of Science and Technology, N-7491, Trondheim, Norway}
\footremember{MAR}{SINTEF Ocean, Marinteknisk senter, N-7052 Trondheim, Norway}
and Yaron Toledo\footrecall{TAU}}

\begin{document}
\maketitle

\abstract{The work analyses the error in current retrievals from images of marine radars. The study is based on simulations of waves interacting with a shear current. The measured dispersion is related to the underlying wavenumber-dependent effective current. The highest tested radar antenna (H=45 m) with vertical polarization performed the best. For that case the root mean square error was at most 0.05 ms$^{-1}$ above the one for the simulated wave field without imaging mechanism. The observation time of 20 minutes was compared to shorter windows. Depending on the needed accuracy, the time may be reduced to five minutes, associated with an loss of accuracy below 12\%. The study shows the error of the current reconstruction depends on the shape of the profile and varies considerably from realization to realization.}

\section{Introduction}
Currents transport sediments as well as nutrients and pollutants. For studying their distribution in the ocean, current profile along the vertical axis is needed. While it can be measured well at some depth, e.g. by ADCPs, it is difficult to obtain the current profile accurately in the uppermost layer where the water particles are affected by the currents and waves simultaneously \citep{DAVIS1981_part2}. High frequency (HF) radars are commonly used to get current estimates close to the surface. Given that the current has a boundary layer underneath the surface, multiple measurements at different wavelengths would be required to define the profile. Although  \citet{ivonin2004validation} have shown that very high frequency (VHF) radars may be used to retrieve currents at three different depth more points would be necessary for an accurate description of boundary layer. 

Marine X-band radars are capable of observing ocean waves of a range of wavelengths, typically from 30 meters up to hundred meters and more. \citet{Young1985} was among the first to describe how these images could be used to extract information on waves and currents by applying a multidimensional Fourier analysis. 
Although \cite{Young1985} acknowledged the work of \citet{stewart1974hf} resulting in a wavenumber-dependent distortion of the dispersion relation, for many years the radar community only considered a bulk current. In recent years there have been multiple attempts for obtaining more detailed measurements of the current in the upper ocean layer from X-band radar images, amongst others,~\citet{lund2018near} and \citet{campana2016}. In their work current profiles are inverted from radar images and compared to ADCPS measurements. As pointed out by \citet{smeltzer2021current_mapping},  a validation of the current estimates of radar images by ADCPs is questionable. The deviation between the ADCP and the radar estimate is in the order of magnitude of the measured current. A slightly similar shear profile comparison presented in \citet{lund2020marine}. It did not only use ADCPs for comparison but a combination of multiple instruments but the problem persists with a ground truth current given for a time period of 3.5 hours. In addition to the different time scales, the reference systems for both instruments differ. While the Doppler-like shift measured by a radar measures the Lagrangian current \citep{pizzo_lenain_romcke_ellingsen_smeltzer_2023}, ADCPs measure the Eulerian current.

inverting the current from the measured Doppler shifts is an ill posed problem \citep{smeltzer2021current_mapping}. The suggested solutions depend on tunable parameters like the smoothness of the final profile \citet{campana2016}. As will be shown below, the selection valid Doppler estimates of the different wavenumber components has an impact that can not be neglected.

This work aims to quantify errors  in shear current profiles estimates by marine radars. The error results from two sources: 1. The accuracy by which the radar measures the dispersion relation. 2. The reconstruction of the underlying shear current. In this study simulations are used to construct artificial radar images. The a priori knowledge of the exact shear profiles provides the errors resulting from both sources separately. A range of different conditions of the radar settings and sea states are compared. Finally the implications of the measured error on a current inversion are illustrated by employing the inversion method presented by \citet{smeltzer2019improved}.

\section{Simulated radar images of waves on a shear current}\label{sec:obs}
\subsection{Simulated wave fields}
\begin{table}[htp]
\caption{Description of the sea state}
\begin{center}
\begin{tabular}{|c|c|c|}
	Name & Parameter & value\\
	\hline
	Significant wave height 	&	$H_s$		& 2.0 m\\
	Peak wave number			&	$k_p$		& 0.073 rad m$^{-1}$\\
	Peak wave length			&	$\lambda_p$	& 87 m\\
	Peak period 				&	$T_p$		& 7.5 s\\
	Directional spreading\footnote{according to \citet{mitsuyasu1975}}  	&	$s_{\max}$ 	& 10, 30, 70\\
	Spatial grid size			&	$\Delta x = \Delta y$	& 7.5 m\\
	Temporal grid size			&	$d_t$		& 1 s\\
	Covered time period			&	$T$			& 20 min\\
	Covered patch size			&	$X=Y$       & 500 m\\
	Peak enhancement factor		&	$\gamma$	& 3.3\\
	Shear current				&	 $U(z) $ 	& $\exp(mz) + 0.05$\\
	Exponential factor			&	$C$			& 0.5, 0.2\\
	Current direction 			&	$\psi$		& $30^\circ$\\
	Wave direction				&	$\theta$	& $90^\circ$\\
	Water depth				    &	$h$			& 1000 m\\
\end{tabular}
\end{center}
\label{tab:overview}
\end{table}%

Simulated wave fields are based on directional JONSWAP spectra and a shear current according to the description provided in Table~\ref{tab:overview}.
It is assumed that non-linear features of waves have no major effect on the radar images and the wave simulations are purely linear. The directional spreading was implemented following \citet{mitsuyasu1975}. For most investigations the current profile $U(z) = \exp(0.5z) + 0.05$ was employed. Part of the results are also shown for a exponential profile with a milder shear ($U(z) = \exp(0.2z) + 0.05$).

\subsection{Incorporation of the shear current}
The interaction of wave motion and shear currents can be expressed by the Rayleigh equation, here using the formulation by \citet{ellingsen2017approximate}
\begin{equation}
w''(z) - k^2  w(z) =  \frac{\mathbf{k}\cdot\mathbf{U}''(z)}{\mathbf{k}\cdot\mathbf{U}(z)-kc}w(z) 
\end{equation}

where $w$ is the vertical velocity , z, the vertical axis with 0 in the mean water level and the negative axis towards the sea bottom. $c$ denotes the wave celerity and $\mathbf{U}$, the two-dimensional shear current.

The boundary condition at $z=\eta(0)$ is
\begin{equation}
\left( \mathbf{k\cdot U}_0 -kc \right)^2 w'(0) - \left( \mathbf{k\cdot U}_0'\left(\mathbf{k\cdot U}_0-kc\right) + \mathrm{g}k^2\right)w(0)=0,
\end{equation}
At the sea bottom the vertical velocity vanishes $w(-h)=0$.

The celerity can be retrieved by the shooting method and the solution is very sensitive to boundary values at the sea surface.
For a wide range of profiles, it is possible to use an approximate solution following the perturbation approach by \citet{stewart1974hf} or higher order versions, e.g. \citet{kirby1989surface}. To first order, the celerity is that of waves without current. The second order solution adds the effective current,
\begin{equation}
U_{\mathrm{eff}}(k) = 2k\int_{-\infty}^0 U(z) \mathrm{e}^{(2kz)}.
\label{eq:effective_current}
\end{equation}
that can be applied to adapt the dispersion relation
\begin{equation}
\omega = \sqrt{k \mathrm{g} \tanh(kh)} + k U_{\mathrm{eff}}(k) \cos(\phi)
\label{eq:dispersion}
\end{equation}
where $\omega$ is the angular frequency, $k$ the wavenumber, $\mathrm{g}$ the gravitational constant, h the water depth and $\phi=\theta-\psi$ the angle between the current $\psi$ and the wavenumber vector $\theta$ (compare \cite{ivonin2004validation}).

\citet{ellingsen2017approximate} introduced the dimensionless depth-averaged shear
\begin{equation}
\delta(\mathbf{k}) = \int_{-h}^0\frac{\mathbf{k\cdot U}'(z) \sinh\left(2k(z+h)\right)}{kc_0 \sinh(2kh)}dz
\end{equation}
that can be used as measure to judge the applicability of the effective current according to \citet{stewart1974hf}.
In all simulations used $|\delta(\mathbf{k})|<0.08$. The largest $\delta$ for the presented cases occurs for the shear current $U(z) = \exp(0.5z)+0.05$. The approximated $\omega(k)$ based on the effective current of \citet{stewart1974hf} deviates at most by 0.0022 $\mathrm{rad s}^{-1}$ from the solution of the Rayleigh equation. The error from the approximation is hence acceptable.

\subsection{Simulated radar images}

The radar image is based on a simplified version of the Radar equation (e.g. \citet{Rees2013})
\begin{equation}
P_r = \frac{P_tG_t\sigma A_e}{(4\pi)^2r^4}
\label{eq:radar_equation}
\end{equation}
where $P_r$ and $P_t$ are the received and transmitted power respectively, $G_t$ is the gain, $A_e$ the effective area (only depending on the radar), $\sigma$ the radar cross section and $r$ the range. If an uncalibrated radar is used and we do not take into consideration the variation in areal coverage we are left with 
\begin{equation}
P_r \propto \frac{\sigma^0}{r^4}
\label{eq:simplified_radar_equation}
\end{equation}
According to \cite{Rees2013}, Eq. 3.51, the backscattering coefficient $\sigma^0$ for the different polarizations $pp$ is
\begin{equation}
\sigma_{pp}^0 = 4 k^4L^2(\Delta h)^2 \cos^4(\theta_l) \left| f_{pp}(\theta)\right|^2 \exp\left(-k^2L^2 \sin^2(\theta_l)\right)
\label{eq:sig0}
\end{equation}
where $k$ denotes the wavenumber of the electromagnetic wave while $\Delta h$ and $L$ denote characteristic scales to define the surface roughness. Herein, it is assumed that these will be constant for a given sea state. Hence the backscattering coefficient only depends on the local incidence angle $\theta_l$. The latter is defined as the angle between the surface normal and the beam towards the radar (given below).
\begin{align}
f_{HH}(\theta) &= \frac{\cos(\theta_l) - \sqrt{\varepsilon_r - \sin^2(\theta_l)}}{\cos(\theta_l) + \sqrt{\varepsilon_r - \sin^2(\theta_l)}},\\
f_{VV}(\theta) &=(\varepsilon_r-1) \frac{\sin^2(\theta_l) - \varepsilon_r  \left( 1 + \sin^2(\theta_l)\right)}{\left(\varepsilon_r \cos(\theta_l) + \sqrt{\varepsilon_r - \sin^2(\theta_l)}\right)^2},
\end{align}
and $k$ is the wave number of the electromagnetic wave, $\Delta h$ the height variation of the roughness and $L$ the correlation length of the roughness of the observed surface.

Depending on the water temperature, $\varepsilon_r$ is in the range of $[70^\circ;80^\circ]$. For high grazing angles, 
\begin{align}
\left|f_{HH}\right|\approx 1
\end{align}
and hence $\sigma_{pp}^0 \propto \cos^4(\theta_l)$. For vertical polarization the local incidence angle changes the backscattering coefficient. The above equation may be approximated by the following:
\begin{align}
\left|f_{VV}\right| \approx \left|\frac{-1-\sin^2(\theta_l)}{\cos(\theta_l)}\right|
\end{align}

The final simplistic radar image is hence given by 
\begin{equation}
P_r \propto \left\{
\begin{matrix}
\frac{\cos^4(\theta_l)}{r^4} & \mbox{for~HH}  \\
\frac{\cos^2(\theta_l)(-1-\sin^2(\theta_l))^2}{r^4} & \mbox{for~VV}
\end{matrix}
\right.
\end{equation}
The amplitude of the signal is the square root of the received power.

The local incidence angle is calculated from the tilt of the surface:
\begin{equation}
\cos\left(  \theta_l \right) = \frac{ r\frac{\partial \eta}{\partial x} + H - \eta }{ \sqrt{1 + \left(\frac{\partial\eta}{\partial x}\right)^2+ \left(\frac{\partial\eta}{\partial y}\right)^2} \sqrt{r^2 + \left(H-\eta\right)^2}} 
\label{eq:Tilt}
\end{equation}
In addition, geometric shadowing as formulated in \citet{Nieto2004} is applied.

\section{Extraction of wavenumber dependent Doppler shift}
We assume the linear dispersion relation for wave frequency $\omega_{\mathrm{DR}}$ as a function of wavenumber $\mathbf{k}$ may be expressed as: 
\begin{equation}
\omega_{\mathrm{DR}}(\mathbf{k}) = \omega_0(k) + \mathbf{k}\cdot\mathbf{U}_{\mathrm{eff}}(k),
\label{eq:DRshear}
\end{equation}
where the $\omega_0(k)$ is the dispersion relation in quiescent waters, $k = |\mathbf{k}|$, $\mathbf{U}_{\mathrm{eff}}(k)$ is the wavenumber-dependent Doppler shift velocity due to the background current profile \cite{stewart1974hf}.

The input data is a three-dimensional array or ``cube'' of sea surface images as a function of time, denoted as $I(x,y,t)$, where $x,y$ correspond to the spatial dimensions of the images, and $t$ is time. The physical extent of the images (assumed to be square) is length $\Delta x$, and the duration of the time series $T$. The images may in this context refer to either optical imagery obtained from an aerial platform for example, or radar backscatter images. The first step in extracting the Doppler shift velocities involves computing the frequency-wavenumber spectrum from a fast Fourier transform:

\begin{equation}
P(\mathbf{k}, \omega) = |\mathrm{FFT}\{I(x,y,t)\}|^2.
\label{eq:spec}
\end{equation}
The spectrum is peaked along the dispersion relation shell. Due to the symmetries of the Fourier transform of a signal comprised of real numbers, there are two wave frequencies $\omega_\pm$ associated with each unique wavenumber: $\omega_+(\mathbf{k}) = \omega_{\mathrm{DR}}(\mathbf{k})$, and $\omega_-(\mathbf{k}) = -\omega_{\mathrm{DR}}(-\mathbf{k})$.

In the following steps, we wish to extract $\mathbf{U}_{\mathrm{eff}}(k)$ from the measured spectrum $P$ over a range of wavenumbers. To do so, we consider wavenumber bins with center values $k_i$ and half-width $\delta k$. A lower limit on $k_i$ of 6-8 multiples of the wavenumber resolution $2\pi/\Delta x$ is typically chosen. On the high wavenumber end, the signal-to-noise ratio typically limits the Doppler shift extraction, and some examination of the data is required in choosing the upper wavenumber limit. For each center wavenumber value $k_i$, we define a masked spectrum $F_i$ where energy outside the wavenumber bin is set to zero, i.e.
\begin{equation}
F_i(\mathbf{k}, \omega) = 
\begin{cases}
\sqrt{P(\mathbf{k}, \omega)}, & \text{if } |k-k_i|\leq \delta k\\
0,   & \text{otherwise}
\end{cases}.
\label{eq:g}
\end{equation}

We define a characteristic function as
\begin{equation}
G(\mathbf{k},\omega,\mathbf{U}_{\mathrm{eff}}) = \mathrm{exp}\left[-2\left(\frac{\omega-\omega_{\mathrm{DR}}(\mathbf{k},\mathbf{U}_{\mathrm{eff}})}{a}\right)^2\right] + \mathrm{exp}\left[-2\left(\frac{\omega+\omega_{\mathrm{DR}}(-\mathbf{k},\mathbf{U}_{\mathrm{eff}})}{a}\right)^2\right],
\label{eq:cf}
\end{equation}
where $a$ is a parameter controlling the spectral width in frequency of the dispersion shell, and we have included $\mathbf{U}_{\mathrm{eff}}$ as independent variable affecting the dispersion relation for $\omega_{\mathrm{DR}}$. The two terms on the right hand side reflect the contributions from $\omega_\pm$ respectively. Previous implementations in the literature have used a different form of (\ref{eq:cf}) where $G$ is equal to unity within a narrow frequency interval about the dispersion relation shell, and equal to zero elsewhere \citep{serafino2009novel,huang2012surface,Huang2016}.
The scalar product is defined as:
\begin{equation}
    \langle A(\mathbf{k},\omega),B(\mathbf{k},\omega)\rangle \equiv \int_{-\infty}^\infty\int_{-\infty}^\infty A(\mathbf{k},\omega)B(\mathbf{k},\omega)\mathrm{d}\mathbf{k}\mathrm{d}\omega.
\label{eq:sp}
\end{equation}
In practice, as $A,B$ herein correspond to discrete three-dimensional arrays, the integrals in (\ref{eq:sp}) are be evaluated as a summation of element-wise multiplications.

The normalized scalar product between the spectrum and characteristic function $G$ is then:
\begin{equation}
V(\mathbf{U}_{\mathrm{eff}}) = \frac{\langle F_i(\mathbf{k}, \omega),G(\mathbf{k},\omega,\mathbf{U}_{\mathrm{eff}})\rangle}{\sqrt{\langle F_i,F_i\rangle\langle G,G\rangle}}.
\label{eq:nsp}
\end{equation}
Maximizing $V$ with respect to the Doppler shift velocity $\mathbf{U}_{\mathrm{eff}}$ implies the greatest overlap between the dispersion relation shell defined by the characteristic function and that of the true spectrum. 

The maximization is performed in two steps. The first involves evaluating $V$ over a wide search grid range covering the expected possible values of $\mathbf{U}_{\mathrm{eff}}$. The value of the current with maximum $V$ over this first search range is $\mathbf{U}_{\mathrm{eff}}^{\mathrm{initial}}$. The second step further maximizes $V$ using $\mathbf{U}_{\mathrm{eff}}^{\mathrm{initial}}$ as an initial guess to a function maximizing routine, which finds a more precise value $\mathbf{U}_{\mathrm{eff}}^{\mathrm{final}}$. In a MATLAB implementation, the solver \textit{fminsearch} is used to minimize $-V(\mathbf{U}_{\mathrm{eff}})$.

\section{Results}\label{sec:results}
\subsection{Current estimates from simulated radar images and simulated waves}
\begin{figure}
\centering
\includegraphics[width=\textwidth]{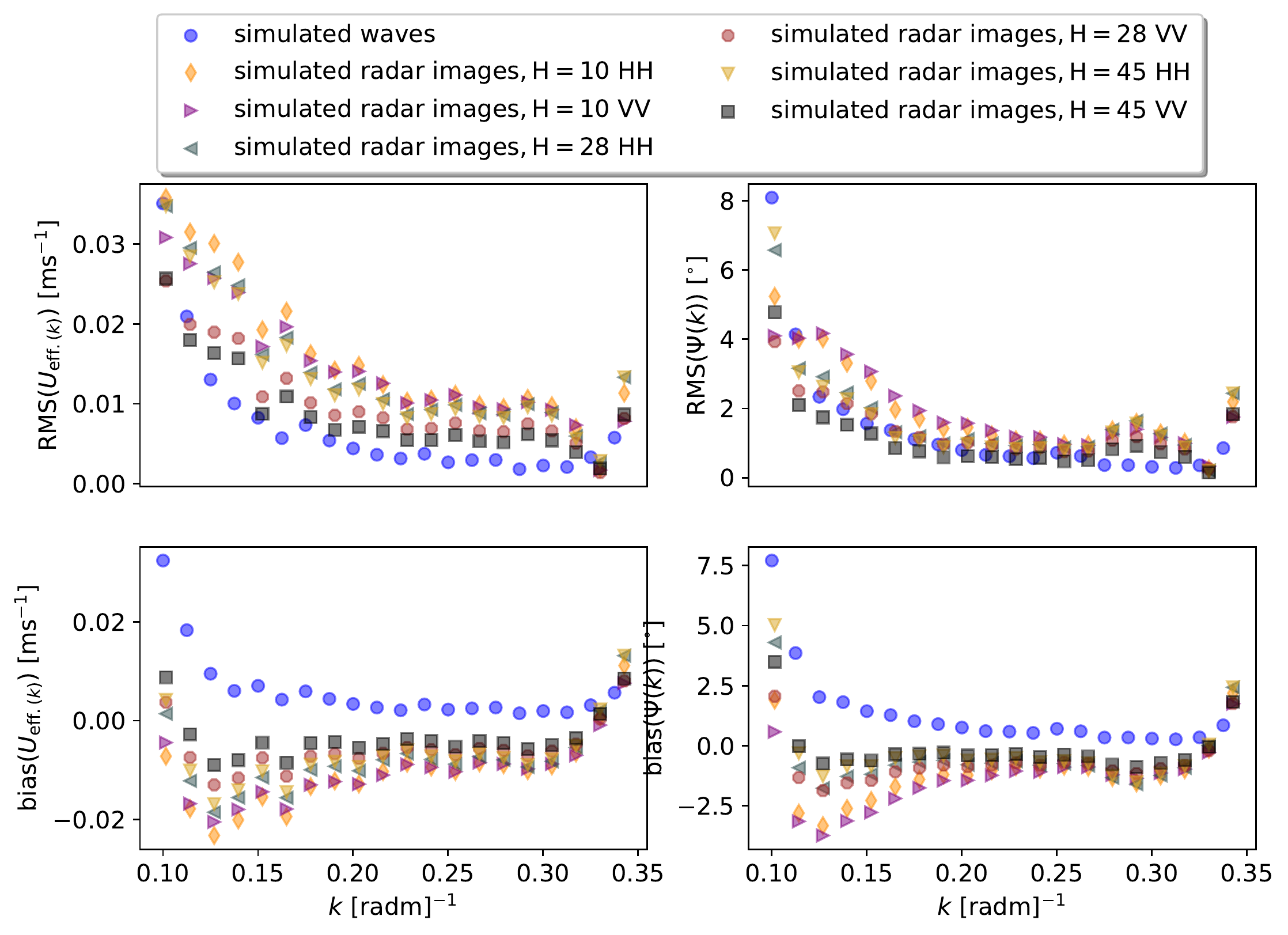}
\caption{Root mean square error and bias of the effective current (left) and direction (right)  based on 50 Simulations. Simulated wave field and 6 different radar settings for the current $U(z)=\exp(0.5z)+0.05$.}
\label{fig:ueff_realizations1}
\end{figure}
\begin{figure}
\centering
\includegraphics[width=\textwidth]{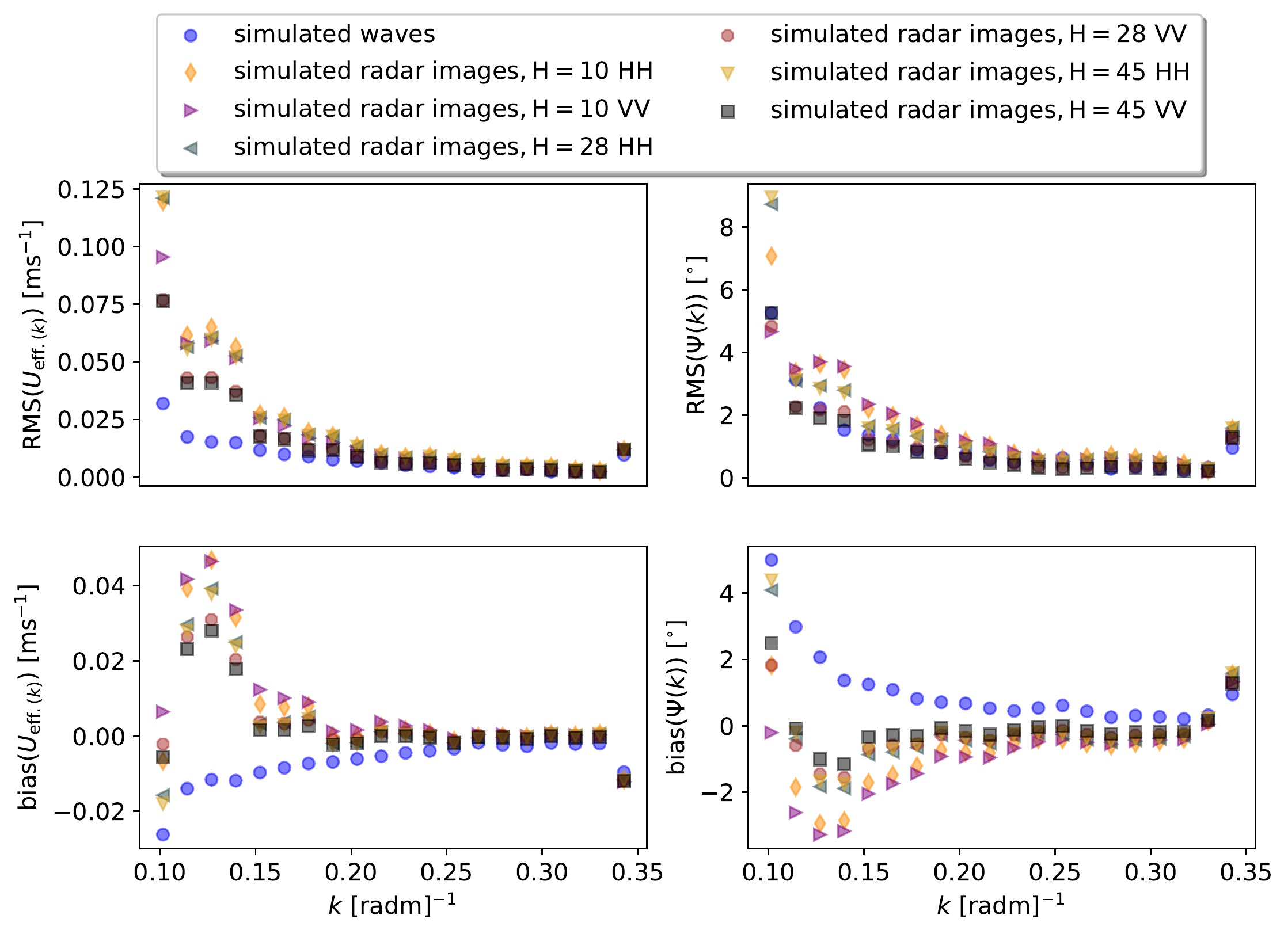}
\caption{Root mean square error and bias of the effective current and direction  based on 50 Simulations. Simulated wave field and 6 different radar settings for the current$U(z)=\exp(0.2z)+0.05$.}
\label{fig:ueff_realizations2}
\end{figure}
The effective current is best estimated for the simulated wave field (compare figures~\ref{fig:ueff_realizations1}~and~\ref{fig:ueff_realizations2}). The simulated radar images deviate with different magnitude depending on the elevation of the radar antenna and the polarization. A height of 45 m and horizontal polarization give the best results with a maximum RMS deviation of about 0.05 m/s. With these radar conditions the estimate of the current direction outperforms even the pure simulated waves up to a wavenumber of 0.27 rad/m. The bias has a different sign for simulated waves and simulated radar images. The effect of the bias for the current inversion will be discussed below. It is worth noticing that the bias for the simulated waves and the simulated radar images have opposite signs. The bias has an order of magnitude similar to the RMS error. Hence, the imaging mechanism systematically shifts the dispersion relation depending on the elevation of the antenna height and the polarization.
The results for the simulations for a current with less shear in the upper layer ($U(z)=\exp(0.2z)+0.05$) were compared.

\subsubsection{Wave directional bandwidth and wave-current angle}
\begin{figure}
\centering
\includegraphics[width=\textwidth]{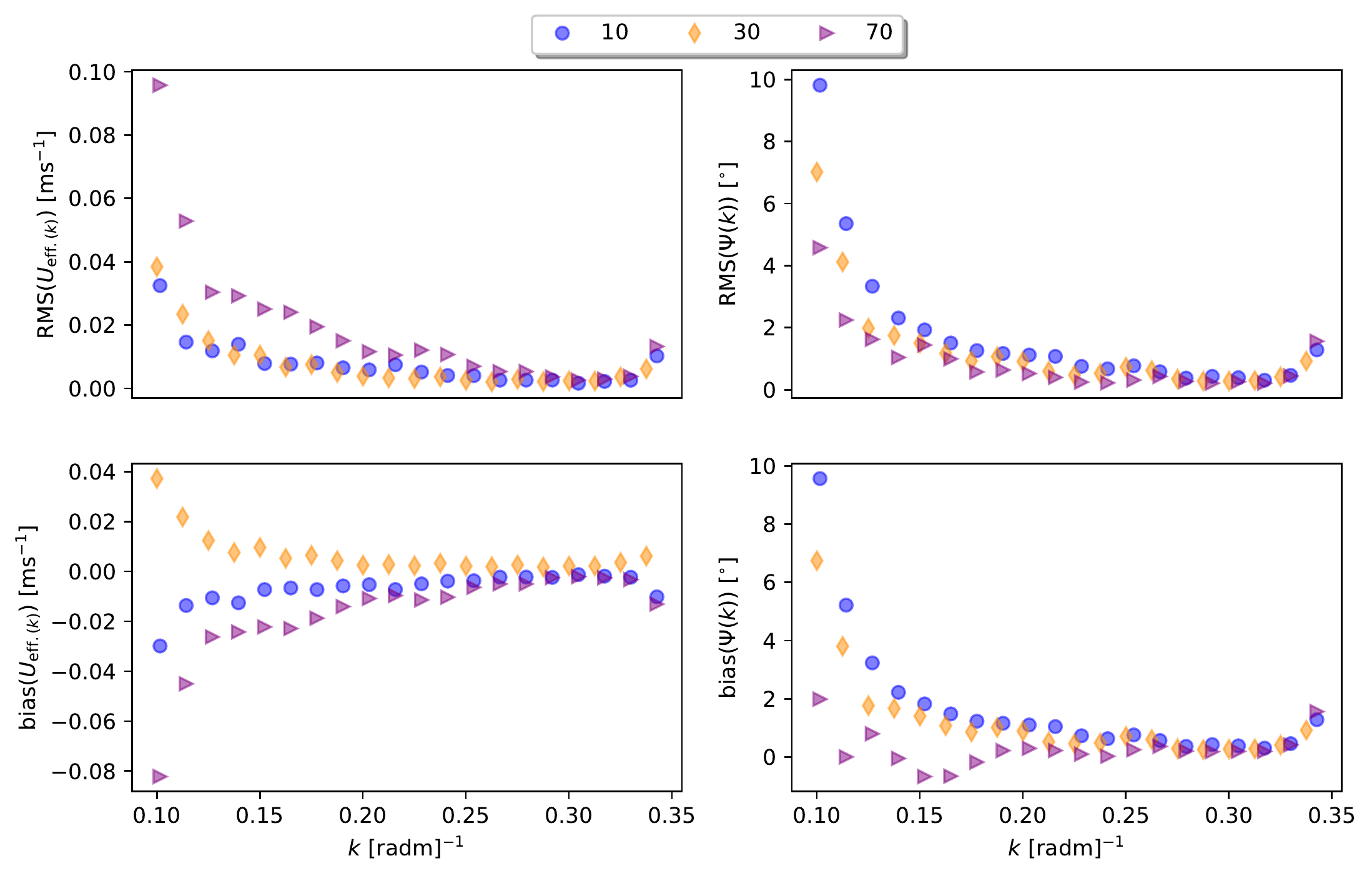}
\caption{The influence of the directional spreading for $\mathrm{s}_{\max}=10,30,70$ (following \citet{mitsuyasu1975} where large spreading is associated with a low value). Statistics based on 10 simulated wave fields for each directional spreading.}
\label{fig:ueff_smax}
\end{figure}
In an ideal setting for remote sensing of currents, the wave field would consist of waves traveling in all directions. Realistic spectra are finite in directional bandwidth however, and in this section we analyze the effect of the directional bandwidth on the extracted Doppler shifts. The Doppler shifts of the values $s_{\mathrm{max}}=10, 30, 70$ were compared for 10 cases of simulated wave fields each (compare Fig.~\ref{fig:ueff_smax}). As the directional bandwidth increases, the Doppler shifts is improved especially in the lower wavenumbers. It should be noted that the bias changes sign for the estimate of the effective current for different cases.

\begin{figure}
\centering
\includegraphics[width=\textwidth]{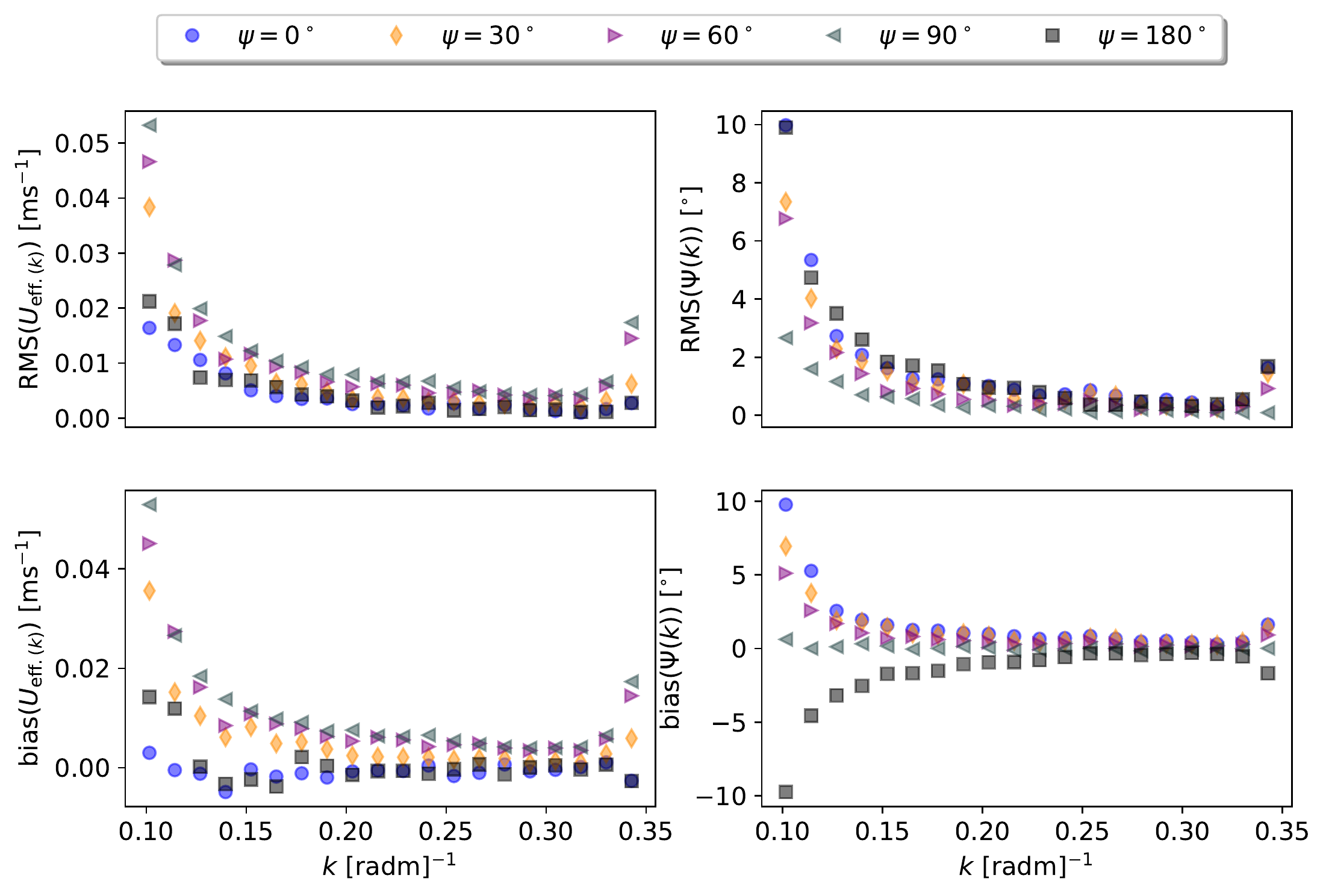}
\caption{Root mean square error and bias for different angles between simulated waves and current (magnitude:left and angle: right).}
\label{fig:angle_comparison}
\end{figure}
The angle between the wave system and the current has a stronger effect on the results than the configurations of the radar system in the low wavenumber regime (compare Figure~\ref{fig:angle_comparison}). For higher wavenumbers, the difference is still noticable. The more current and waves are parallel, the lower the error of the magnitude of the current. For the angle, it is opposite but the differences are marginal.

\subsubsection{Temporal and spatial window size}
\begin{figure}
\centering
\includegraphics[ width=\textwidth]{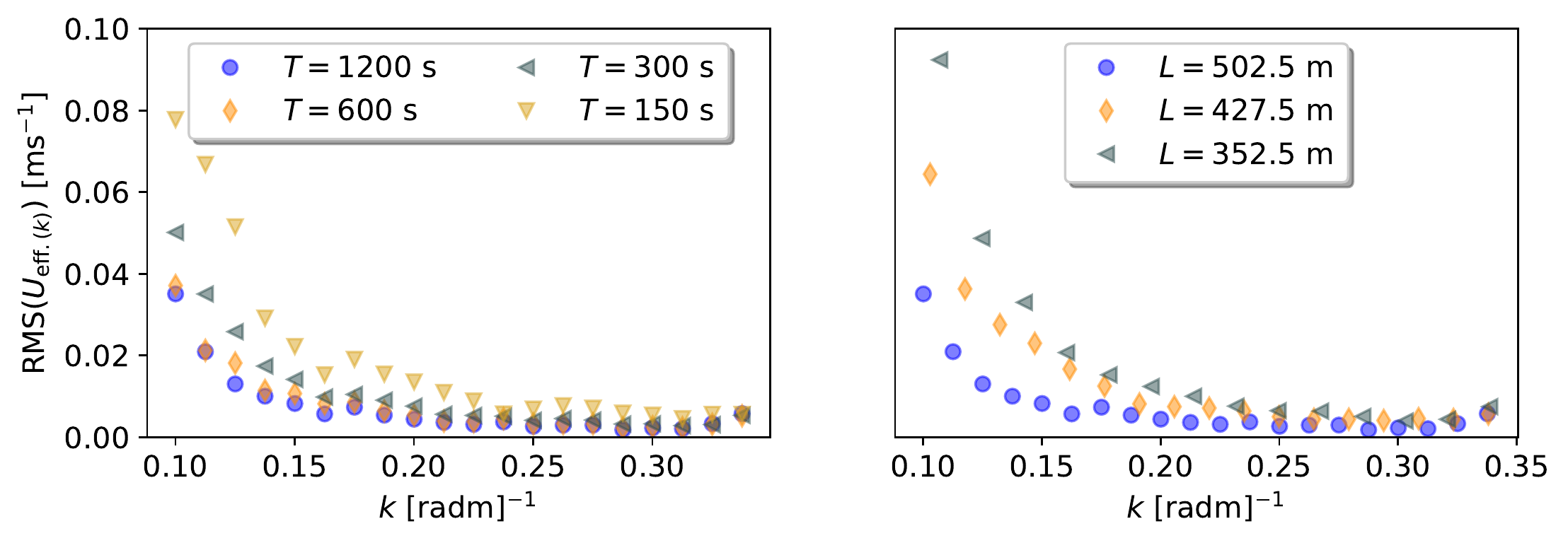}
\caption{Dispersion retrieval for different temporal and spatial windows.}
\label{fig:ueff_T_L}
\end{figure}

The finite dimensions of the sea surface image cube result in corresponding spectral resolutions $\Delta\omega$ and $\Delta k$. The spectral resolution in turn limits the accuracy with which the dispersion relation and Doppler shift velocities can be determined. In this section we explore the effect of the temporal and spatial window sizes on the accuracy of the Doppler shifts. In figure \ref{fig:ueff_T_L} we show the Doppler shift velocities as a function of wavenumber for different temporal durations $T$ and window lengths $L$ as indicated in the respective legends. The results are not particularly sensitive to $T$ for values greater than $300 s$ (five minutes). 
The governing spectral resolution is imposed by the spatial window which should not be reduced below five peak wave lengths. 
The sensitivity to the window size $L$ is strong, especially for larger wavenumbers. The window size limits the minimum wavenumbers where Doppler shift velocities may be extracted, thus having consequences for the range of depths for which the current profile can be reconstructed.

To interpret the results of Fig.~\ref{fig:ueff_T_L}, we estimate the uncertainty scalings of the dispersion relation due to the spectral resolution \citep{smeltzer2019improved}:

\begin{equation}
    \Delta c_{\Delta k} = \frac{\mathrm{d}\omega_0}{\mathrm{d}k}\Delta k / k
\end{equation}

\begin{equation}
    \Delta c_{\Delta \omega} = \Delta\omega / k.
\end{equation}

\begin{figure}
\centering
\includegraphics[trim = {0cm 0 0cm 0},clip, width=.5\textwidth]{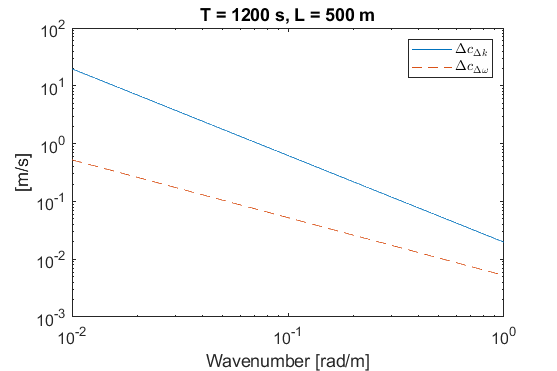}
\caption{Uncertainty scaling of the Doppler shift velocities based on spectral resolution $\Delta k$ and $\Delta\omega$.}
\label{fig:unc}
\end{figure}

The parameters $\Delta c_{\Delta k}$ and $\Delta c_{\Delta \omega}$ may be interpreted as the spectral resolution in units of velocity. We show in figure \ref{fig:unc} $\Delta c_{\Delta k}$ and $\Delta c_{\Delta \omega}$ for $T = 1200$ s and $\Delta x = 500$ m. Over all relevant wavenumbers, $\Delta c_{\Delta k}$ is greater than $\Delta c_{\Delta \omega}$, meaning that uncertainty in the dispersion relation is dominated by the wavenumber resolution as opposed to frequency resolution. We now can interpret the results shown in Fig.~\ref{fig:ueff_T_L}: the errors in the Doppler shift velocities are more sensitive to decreasing $\Delta x$ due to the increase in $\Delta c_{\Delta k}$, whereas decreasing $T$ up to point has little effect as $\Delta c_{\Delta k}$ still is dominant. It is thus only when $T$ is decreased such that $\Delta c_{\Delta\omega}\sim\Delta c_{\Delta k}$ that the uncertainty of the dispersion relation becomes affected.

\section{Profile reconstruction}\label{sec:profile_reconstruction}
This section evaluates how the error from the measurement of the dispersion relation may affect the reconstruction of the current profile. Multiple reconstruction methods have been suggested. A summary may be found in \citet{smeltzer2021current_mapping}. Details on the inversion method employed herein can be found in \citet{smeltzer2019improved}. It is not the purpose of this work to provide the best current reconstructions but to illustrate how errors in the measured dispersion relation change the result of the inversion.

In the course of this work it has been noticed that the shear profile with the stronger shear studied above is not well suited for the reconstruction of the current. Even when employing input of the correct dispersion relation. The analysis of the effective current is therefore restricted to the case with lower shear, $U(z)=\exp(0.2z)+0.05$.

 The result for low wavenumbers is strongly affected by spectral leakage.

\begin{figure}
    \centering
    \includegraphics[width=\textwidth]{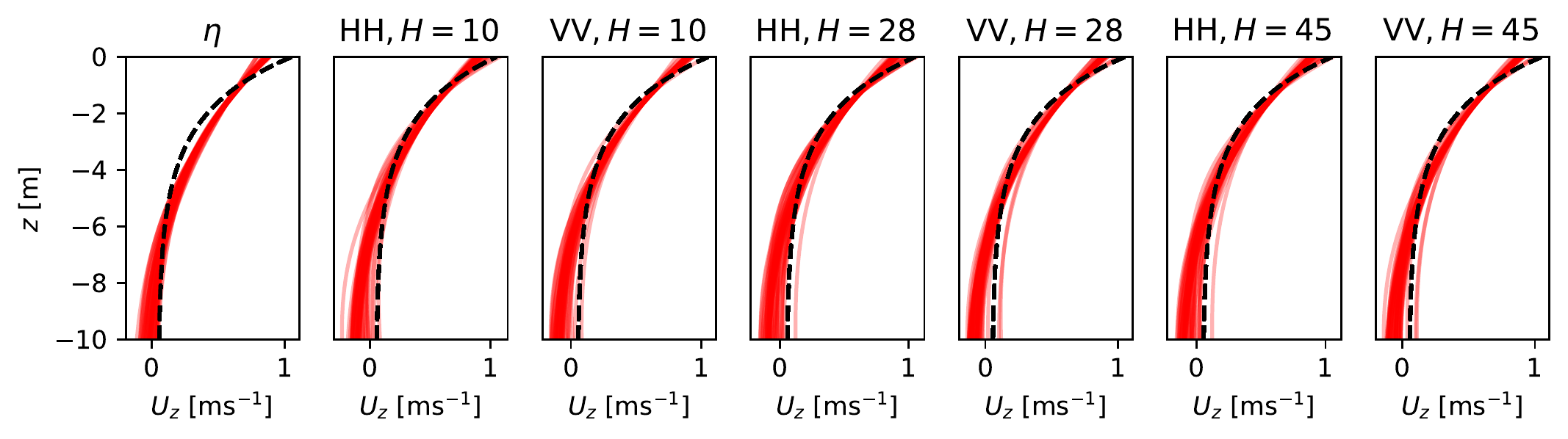}
    \includegraphics[width=\textwidth]{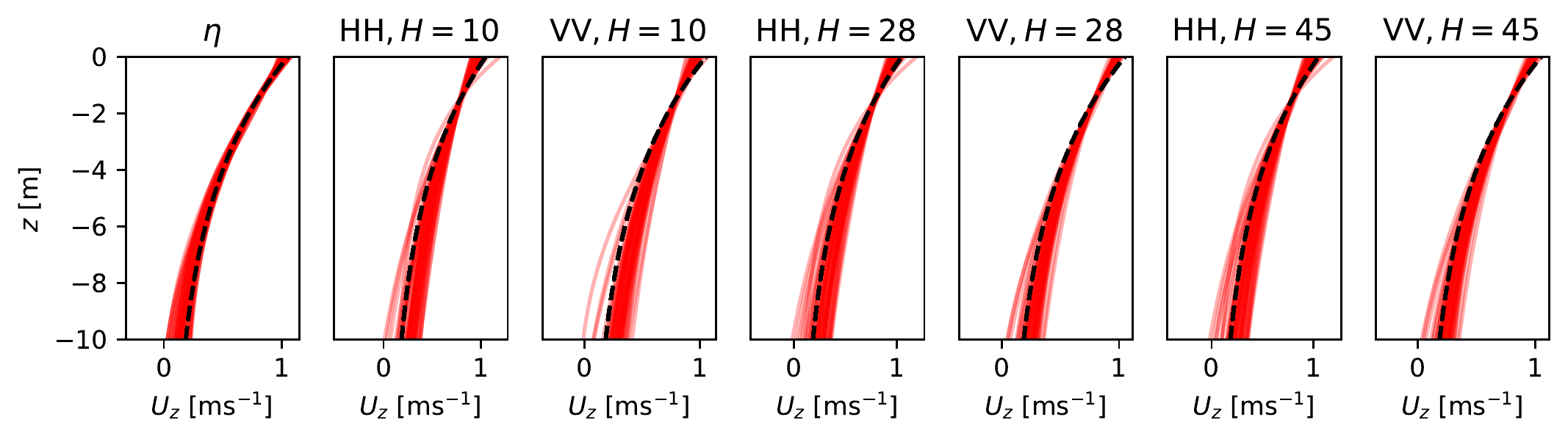}
    \caption{Current inversions for simulated waves ($\eta$) and simulated radar images (polarization and antenna height as provided in the title) for 50 examples each. The true shear current is drawn as black dashed line. It is $U(z)=\exp(0.5z)+0.05$ in the upper row and $U(z)=\exp(0.2z)+0.05$ in the lower row. }
    \label{fig:PEDM}
\end{figure}

\begin{figure}
    \centering
\includegraphics[width=0.49\textwidth]{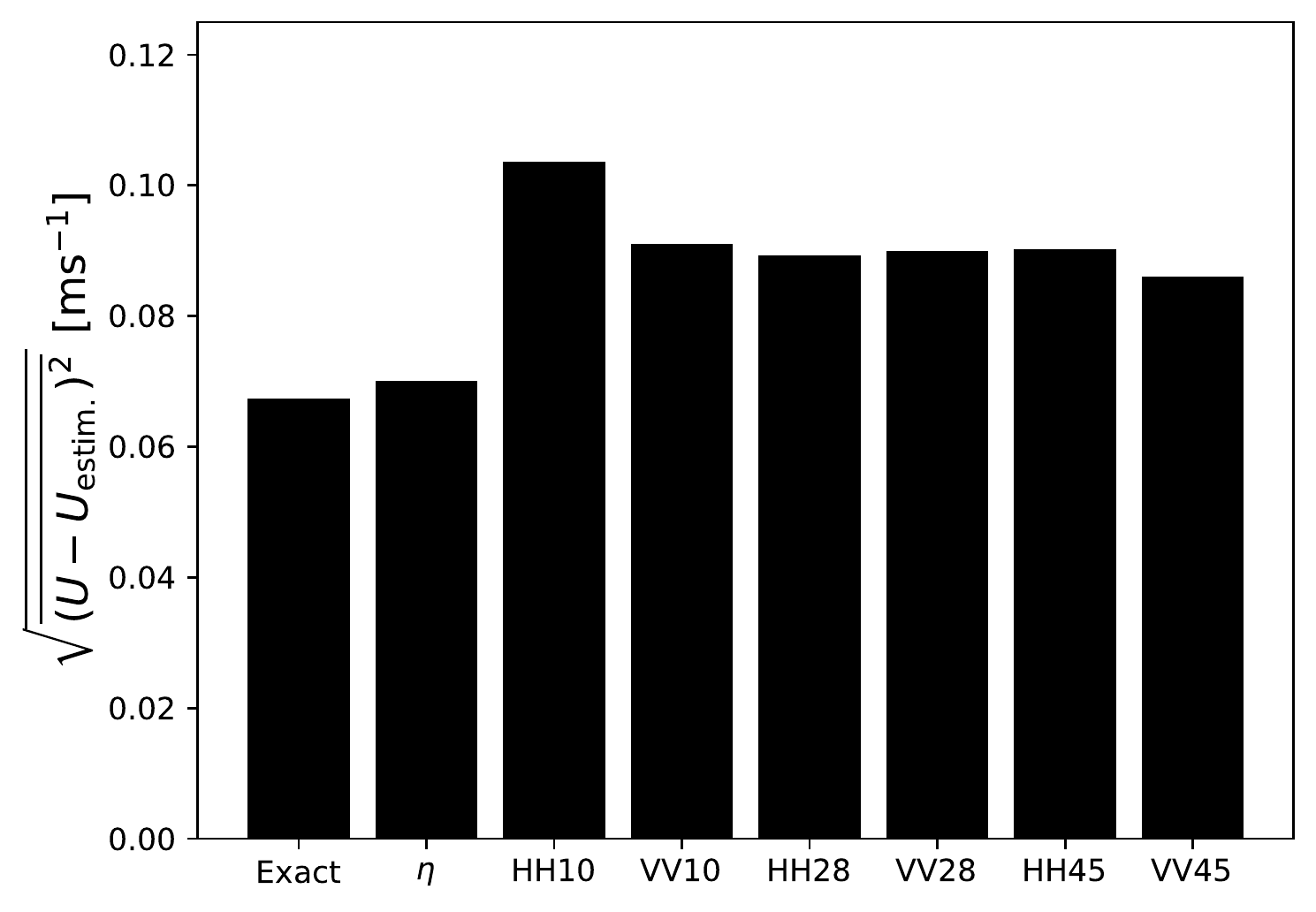}
    \includegraphics[width=0.49\textwidth]{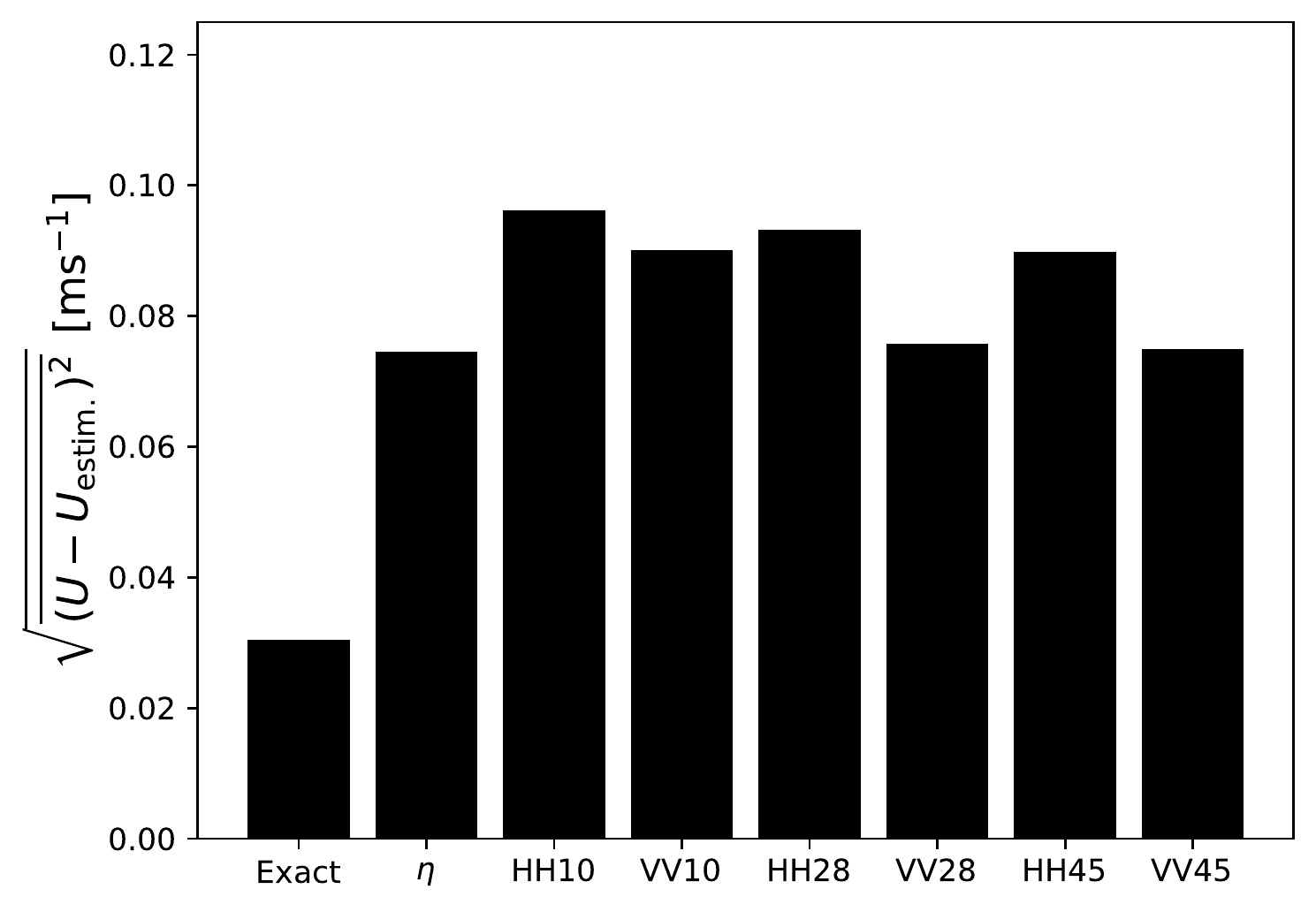}
    \caption{Overall performance of current inversions for $U(z)=\exp(0.5z)+0.05$ (left)  and $U(z)=\exp(0.2z)+0.05$ (right) as shown in Fig.~\ref{fig:PEDM}. }
    \label{fig:RMS_PEDM}
\end{figure}

The shear currents are retrieved by the PEDM algorithm \citep{smeltzer2019improved}. Each realization is represented by single line in Fig.~\ref{fig:PEDM}. Superimposed lines result in a a darker colour in the plot. The true profiles are shown as dashed black lines. Due to the strong bias that some of the dispersion measurements have shown, a pre-processing step was needed. The effective current is a smooth quantity and sudden jumps in the data can easily be detected and eliminated. Further, the sampling variability leads to a non-smooth curve. By fitting the remaining points to a polynomial of degree five reduces the unwanted noisiness in the measured dispersion relation and improves the shear current reconstruction.

For each case 50 realizations were analyzed. The shear is underestimated in the uppermost layer. In deeper layers the estimated shear current has an offset where the direction of the offset depends on the profile. For very strong shear the method tends to overshoot in lower layers.  Overall, the results for the simulated waves and the different radar configurations lead to similar results. The differences become apparent by evaluating the root mean square (RMS) error as shown in Fig.~\ref{fig:RMS_PEDM}. The first bar shows the error when using the correct effective current as input to the method. It shows that the method is not able to obtain the correct profile even with the correct input, the error also depends strongly on the shape of the current profile. The method is better suited for cases of milder shear as the RMS error for exact input can be reduced by one half with respect to a stronger current. For the Doppler shifts extracted from simulated waves the error is in the range of the exact input or above. Adding the imaging mechanism leads to a wider spreading of solutions for the different realizations. For the two example profiles, this effect is stronger in the profile with the lesser shear.  The increase in spread is more pronounced for horizontal polarization. The 95\% confidence interval ranges up to 3 cm/s up to a water depth of 10 m. The size of the confidence intervals varies up to 1 cm/s between the cases. Overall, the antenna height of $H=45$ m is favourable and vertical polarization should be used.

\section{Discussion and conclusion}\label{sec:discussion}
The presented work confirms that X-band radars are capable of measuring a part of the dispersion relation corresponding to wavenumbers in the range from 0.1 to 0.35 rad/m. By the offset from the theoretical dispersion relation the underlying effective current can be estimated. Depending on the current profile and the wavenumber the RMS lies below 0.1 m/s and below 0.04 m/s for $k>0.15$rad/m. The reconstruction of the shear current is less accurate. Although the average performance over 50 realizations shows reasonable results, there is a considerable spread of single realizations. It has been shown by simulations that the imaging mechanisms amplifies both the deviations from the true profiles, and the spread for different realizations. When comparing the performance of the inversion method by inserting the exact effective current, the error depends strongly on the input profile. For the studied examples the error was doubled. 

It should be noted that the study implements the radar measurement as a Eulerian measurement rather than a Lagrangian one (compare \citep{pizzo_lenain_romcke_ellingsen_smeltzer_2023}). Therefore the Stokes drift that is normally incorporated in current retrievals from radar images is not contained in the simulations. To our understanding it does not affect the findings herein as the Stokes drift is neither present in the wave model nor in the radar image.

Further improvements of the complete scheme may be expected by resolving the dispersion by instruments of higher resolution. As the shear has the strongest effect in the uppermost ocean layer more information in the short wave segment would improve the results. The observed decrease of the error for increasing wavenumbers results from a more evenly distribution of energy in the spectral domain for higher frequencies. For one, the number of waves in relation to the domain is larger and shorter waves have a higher directional spreading.

As outlined above, the spatial resolution limits the range of suitable wavenumbers for the current estimates at the lower end. The influence of the window size may proof challenging in reality. To comply with limitations on the homogeneity of current fields, the patch size was limited to around 500 m in each direction. The wave length was deliberately chosen to not fit into the window. The algorithm of the current retrieval is subject to spectral leakage that causes larger errors. It may help to use wider patches for the lower wavenumbers, assuming that the current in deeper layers is less inhomogeneous. This is however only valid in deep water.

\begin{figure}
    \centering
    \includegraphics[height=0.4\textwidth]{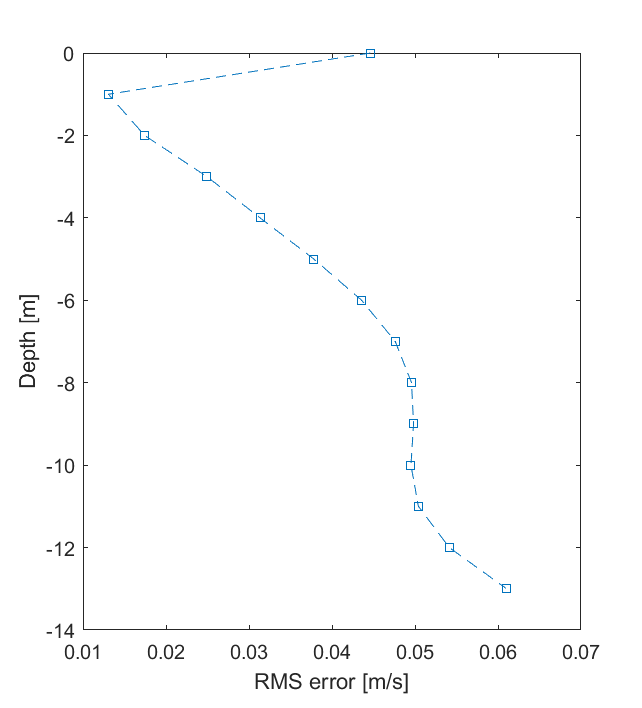}
    \includegraphics[height=0.4\textwidth]{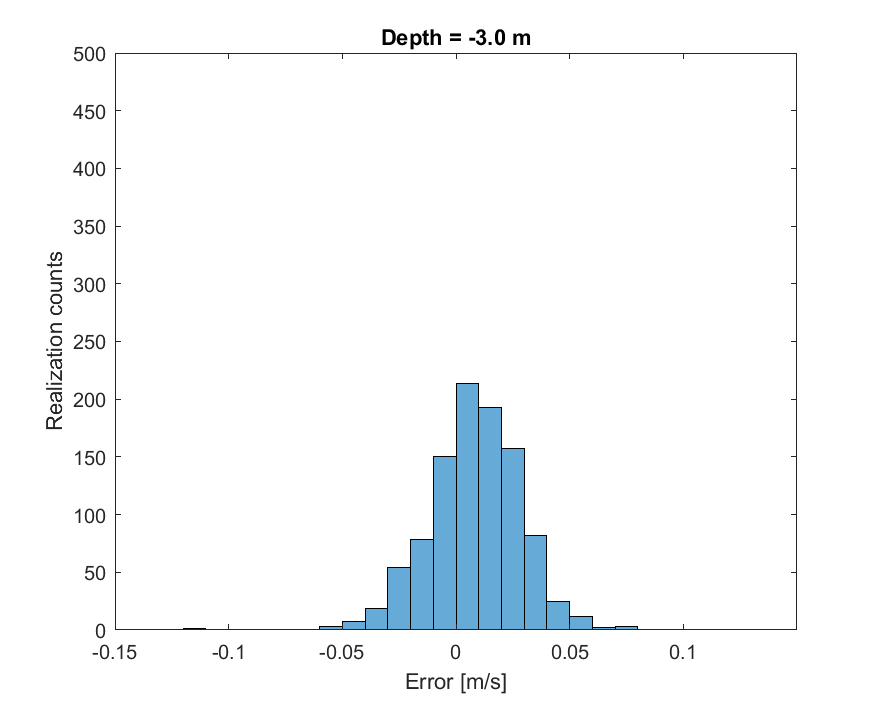}
    \caption{Analysis of influence of random noise on the retrieval of the current. Left: RMS error based on 1000 realizations. Right: Error distribution at three meters below the mean water level.}
    \label{fig:noise}
\end{figure}

Further improvement of the inversion algorithm with a stronger focus on possible noise and biases in the input would be advisable. In the presented study a crucial step was to sort out Doppler estimates in the low wavenumber regime to obtain good shear current reconstructions. Hence, the method is very sensitive to errors. In its current form the simulation of radar images is free of noise and errors are probably underestimated. An additional analysis of the influence of random noise in the current inversions was conducted. 1000 realizations of effective currents ($U(z)=\exp(0.5z)+0.05, \psi=30^\circ$ with random noise were analyzed. It confirms that the errors increase with depth but there is no bias introduced (compare Fig.~\ref{fig:noise}). 

It would be desirable to extend the presented work to intermediate and shallow waters. However, simulating realistic bathymetries with waves and currents is not straightforward. The inhomogenity of condtions cause a smearing of the dispersion cone.

For future campaigns on retrieval of the current based on radar images, a high antenna (e.g. 45 m) and vertical polarization are preferable (for a distance of 200 meters between the radar and the observed waves). If the evolution of the current is of interest 5-10 minute time windows are suggested as trade-off between temporal resolution and low error. The patch size of the observation should include more than five peak wavelengths if homogeneity of the current can be assumed.

\section*{Acknowledgements}
This work was funded by Minerva Fund and the European Research Council through the HIGHWAVE project (grant no. 833125).

\bibliographystyle{apalike}
\bibliography{AllArticles.bib}
\end{document}